\documentclass[oneside,11pt]{article}

\usepackage[utf8]{inputenc}
\usepackage[T1]{fontenc}

\usepackage{amsthm}
\usepackage{amsmath}
\usepackage{amsfonts}
\usepackage{amssymb}

\usepackage{graphicx} 
\usepackage{float}
\usepackage{booktabs}
\usepackage{multirow}
\usepackage{rotating}
\usepackage{array}
\usepackage{xcolor}
\usepackage{subfig}
\usepackage[ruled]{algorithm2e}

\usepackage{breakcites}

\usepackage{enumitem}

\usepackage[top=1.5cm,bottom=2cm,right=2.5cm,left=2.5cm]{geometry}
\usepackage{titling}

\usepackage{natbib}

\usepackage{ifplatform}

\usepackage{textcomp}

\ifwindows
\fi

\ifwindows
\usepackage{bbm}
\else
\iflinux
\usepackage{bbm}
\else
\usepackage{bbm}
\fi
\fi

\newcommand{\lrp}[1]{\left(#1\right)}
\newcommand{\lrc}[1]{\left[#1\right]}
\newcommand{\lrb}[1]{\left\{#1\right\}}
\DeclareFontFamily{OT1}{pzc}{}
\DeclareFontShape{OT1}{pzc}{m}{it}{<-> s * [1.10] pzcmi7t}{}
\DeclareMathAlphabet{\mathpzc}{OT1}{pzc}{m}{it}

\usepackage[pdftex,final,bookmarksnumbered,bookmarksopen=false,breaklinks,colorlinks]{hyperref}
\hypersetup{
	final,
	bookmarksnumbered=true,
	bookmarksopen=true,
	bookmarksopenlevel=0,
	unicode=false,          
	pdftoolbar=true,        
	pdfmenubar=true,        
	pdffitwindow=false,     
	pdftitle={A Cramer-von Mises test of uniformity on the hypersphere},    
	pdfdisplaydoctitle=true, 
	pdftoolbar=true,
	pdfmenubar=true,
	pdflang={English},
	pdfauthor={Eduardo Garcia-Portugues, Paula Navarro-Esteban, Juan A. Cuesta-Albertos},     
	pdfsubject={arXiv paper},   
	pdfcreator={Eduardo Garcia-Portugues},   
	pdfproducer={Eduardo Garcia-Portugues}, 
	pdfkeywords={Circular data}{Craters}{Directional data}, 
	pdfnewwindow=true,      
	breaklinks=true,
	hidelinks,       
	linkcolor=black,          
	citecolor=black,        
	filecolor=magenta,      
	urlcolor=cyan           
}

\newtheorem{theo}{Theorem}
\newtheorem{coro}{Corollary}

\allowdisplaybreaks

\newif\ifmain
\maintrue
\ifmain

\else

\fi
\newif\ifsupplement
\supplementtrue
\newif\iffigstabs
\figstabstrue

\begin{document}

\ifmain

\title{A Cram\'er--von Mises test of uniformity on the hypersphere}
\setlength{\droptitle}{-1cm}
\predate{}%
\postdate{}%

\date{}

\author{Eduardo Garc\'ia-Portugu\'es$^{1,3}$, Paula Navarro-Esteban$^{2}$, and Juan A. Cuesta-Albertos$^{2}$}

\footnotetext[1]{Department of Statistics, Carlos III University of Madrid (Spain).}
\footnotetext[2]{Department of Mathematics, Statistics and Computer Science, University of Cantabria (Spain).}
\footnotetext[3]{Corresponding author. e-mail: \href{mailto:edgarcia@est-econ.uc3m.es}{edgarcia@est-econ.uc3m.es}.}

\maketitle


\begin{abstract}
Testing uniformity of a sample supported on the hypersphere is one of the first steps when analysing multivariate data for which only the directions (and not the magnitudes) are of interest. In this work, a projection-based Cram\'er--von Mises test of uniformity on the hypersphere is introduced. This test can be regarded as an extension of the well-known Watson test of circular uniformity to the hypersphere. The null asymptotic distribution of the test statistic is obtained and, via numerical experiments, shown to be tractable and practical. A novel study on the uniformity of the distribution of craters on Venus illustrates the usage of the test.
\end{abstract}
\begin{flushleft}
	\small\textbf{Keywords:} Circular data; Craters; Directional data.
\end{flushleft}

\section{Introduction}
\label{sec:1}

Testing uniformity of a sample $\mathbf{X}_1,\ldots,\mathbf{X}_n$ of a random vector $\mathbf{X}$ supported on the hypersphere $\Omega_{q}:=\{\mathbf{x} \in \mathbb{R}^{q+1}: \mathbf{x}' \mathbf{x}=1\}$ of $\mathbb{R}^{q+1}$, with $q\geq1$, is one of the first steps when analysing \textit{directional data}, that is, data supported on $\Omega_{q}$. Directional data arise in many applied disciplines, such as astronomy or biology, and have been the focus of a considerable number of monographs; see, e.g., \cite{Mardia2000} and \cite{Ley2017}. Since the seminal paper by Lord Rayleigh \citep{Rayleigh1919}, and despite its relative concreteness, the century-old topic of testing uniformity on $\Omega_q$ has attracted more than thirty proposals of tests with varying degrees of generality (many are circular- or spherical-specific tests, i.e., they assume $q=1$ or $q=2$); see \cite{Garcia-Portugues:review} for a review on the topic.\\

Testing uniformity on $\Omega_2$ has several applications in astronomy. An instance is the analysis of the presumed uniform orbit distribution of long-period comets originating in the nearly-isotropic Oort cloud \citep{Cuesta-Albertos2009}. Another application is in the analysis of the distribution of crater impacts, a valuable informer on the impactors that create them. For instance, the case study in \cite{Hirata2016} for Rhea attributes the uniform-like distributions of small craters to  the predominance of planet-orbiting impactors caused from returning debris ejected from large crater impacts. Sun-orbiting impactors, on the other hand, tend to be related to non-uniform crater distributions.\\

In this work, we propose yet another test of uniformity on $\Omega_q$. The test is based on projections, it is of a Cram\'er--von Mises nature, and it has the following main appeals: (\textit{i}) applicability to arbitrary dimensions $q\geq1$; (\textit{ii}) consistency against any alternative to uniformity, i.e., \textit{omnibusness}; (\textit{iii}) conceptual neat extension of the well-known \cite{Watson1961} test of \emph{circular} uniformity; (\textit{iv}) known and usable asymptotic distribution; (\textit{v}) computational tractability for the most common dimensions.\\

The contents of the work are organized as follows. Section \ref{sec:back} sets the problem (Section \ref{sec:H0}), reviews a projection-based test of uniformity that motivates this work (Section \ref{sec:CA}), and exposes the projected uniformity distribution (Section \ref{sec:proj}). Section \ref{sec:test} presents the new test of uniformity, providing the genesis of the test statistic (Section \ref{sec:gen}), its $U$-statistic form (Section \ref{sec:Ustat}), and its asymptotic null distribution (Section \ref{sec:asymp}). Numerical experiments given in Section \ref{sec:num} evidence the tractability of the asymptotic distribution and the fast convergence of the test statistic towards it. Finally, Section \ref{sec:venus} investigates whether Venusian craters are uniformly distributed.

\section{Background}
\label{sec:back}

\subsection{Testing uniformity on $\Omega_{q}$}
\label{sec:H0}

Testing the uniformity of a continuous random variable $\mathbf{X}\sim\mathrm{P}$ supported on $\Omega_q$ is a simple goodness-of-fit problem. It is formalized as the testing of
\begin{align}
\mathcal{H}_0: \mathrm{P}=\nu_q\quad\text{vs.}\quad\mathcal{H}_1:\mathrm{P}\neq \nu_q \label{eq:H0}
\end{align}
from a sample $\mathbf{X}_1,\ldots,\mathbf{X}_n$ of independent and identically distributed observations from $\mathrm{P}$, the distribution of $\mathbf{X}$, and where $\nu_q$ denotes the uniform distribution on $\Omega_q$. The probability density function (pdf) of $\nu_q$ assigns density $\omega_q^{-1}$ to any point on $\Omega_q$, where $\omega_q:=2\pi^{\frac{q+1}{2}}\big/\Gamma\big(\tfrac{q+1}{2}\big)$ denotes the surface area of $\Omega_{q}$, $q\geq1$.\\

If $\mathbf{X}\sim \nu_q$, then $\mathbf{X}$ is identically distributed to any rotation of $\mathbf{X}$. This property  suggests that  any proper test for $\mathcal{H}_0$ must be rotation invariant, in the sense that the obtained test decision should remain invariant if we apply the test to any rotation of the sample. Recall also that, since $\mathcal{H}_0$ is actually a simple hypothesis that completely specifies a distribution, Monte Carlo calibration of any test statistic for problem \eqref{eq:H0} is conceptually straightforward (though perhaps computationally~costly).

\subsection{Using projections for assessing uniformity}
\label{sec:CA}

A projection-based test of uniformity on $\Omega_q$ is proposed in \cite{Cuesta-Albertos2009}. This test is based on Corollary 3.2 in \cite{Cuesta-Albertos2007}, from which it is easily deduced that, under some mild regularity conditions, if
\begin{enumerate}
	\item[(\textit{i})] $\mathbf{X}$ and $\mathbf{Y}$ are two $d$-dimensional random vectors  whose distributions are different, 
	
	\item[(\textit{ii})] $\boldsymbol\gamma$ is a random vector independent from $\mathbf{X}$ and $\mathbf{Y}$ with distribution absolutely continuous with respect to the Lebesgue measure, 
\end{enumerate}
then the distributions of the projections of $\mathbf{X}$ and $\mathbf{Y}$ on the one-dimensional subspace generated by $\boldsymbol\gamma$ almost surely differ. \\

Taking into account that the distribution of the projections coincide if $\mathbf{X} \sim \mathbf{Y}$, we have that testing $\mathcal{H}_0$ is almost surely equivalent to testing $\mathcal{H}^{\boldsymbol{\gamma}}_0: \mathbf{X}'\boldsymbol\gamma\sim \Pi_q$, where $\Pi_q$ is the distribution of $\gamma'{\bf U}$ and ${\bf U} \sim \nu_q$ (see Section \ref{sec:proj}).\\

The test by \cite{Cuesta-Albertos2009} proceeds as follows: (\textit{i}) sample $\boldsymbol\gamma \sim \nu_q$; (\textit{ii}) reject $\mathcal{H}_0^{\boldsymbol\gamma}$, and consequently $\mathcal{H}_0$, for large values of the Kolmogorov--Smirnov statistic
\begin{align}
\mathrm{KS}_{n,\boldsymbol\gamma}:=\sup_{-1\leq x \leq 1} |F_{n,\boldsymbol{\gamma}}(x)- F_q(x)|,\label{eq:KSn}
\end{align}
where $F_{n,\boldsymbol{\gamma}}$ is the empirical cdf of $\mathbf{X}_1'\boldsymbol\gamma,\ldots,\mathbf{X}_n'\boldsymbol\gamma$ and $F_q$ is the cdf of $\Pi_q$.\\

The test that rejects $\mathcal{H}_0$ for large values of $\mathrm{KS}_{n,\boldsymbol\gamma}$ is omnibus and fast to evaluate. However, it is also dependent on $\boldsymbol{\gamma}$, whose selection adds an extra layer of randomness. As proposed in \cite{Cuesta-Albertos2009}, this can be mitigated by considering $k$ random directions and combining the $p$-values associated to each of the $k$ tests into the test statistic
\begin{align}
\mathrm{CCF}_{n,\boldsymbol{\gamma}_1,\ldots,\boldsymbol{\gamma}_k}:=\min\{p\text{-value}_1,\ldots,p\text{-value}_k\},\label{eq:CCFn}
\end{align}
which rejects $\mathcal{H}_0$ for small values. The asymptotic distribution of \eqref{eq:CCFn} is unknown and has to be calibrated by Monte Carlo (conditionally on the choice of $\boldsymbol{\gamma}_1,\ldots,\boldsymbol{\gamma}_k$).

\subsection{Projected uniform distribution}
\label{sec:proj}

The distribution $\Pi_q$ is fundamental to any projection-based test of uniformity. It does not depend on $\boldsymbol{\gamma}$ and its pdf (see, e.g., \citet[page 167]{Mardia2000}) is
\begin{align*}
\mathrm{B}\left(\tfrac{1}{2},\tfrac{q}{2}\right)^{-1}(1-t^2)^{q/2-1}, \quad t\in[-1,1],
\end{align*}
where $\mathrm{B}(a,b):=\Gamma(a)\Gamma(b)/\Gamma(a+b)$. Therefore, $(\gamma'{\bf U})^2\sim\mathrm{Beta}\left(\tfrac{1}{2},\tfrac{q}{2}\right)$ and
\begin{align*}
F_q(x)=\mathrm{B}\left(\tfrac{1}{2},\tfrac{q}{2}\right)^{-1}\int_{-1}^x(1-t^2)^{q/2-1}\,\mathrm{d}t=\frac{1}{2}\left\{1+\mathrm{sign}(x)\mathrm{I}_{x^2}\left(\tfrac{1}{2},\tfrac{q}{2}\right)\right\},
\end{align*}
where $\mathrm{I}_x(a,b):=\mathrm{B}(a,b)^{-1}\int_{0}^xt^{a-1}(1-t)^{b-1}\,\mathrm{d}t$, $a,b>0$, is the regularized incomplete beta function. Trivially, $F_{1}(x)=1-\frac{\cos^{-1}(x)}{\pi}$ and $F_{2}(x)=\frac{x+1}{2}$ for $x\in[-1,1]$.

\section{A new test of uniformity}
\label{sec:test}

\subsection{Genesis of the test statistic}
\label{sec:gen}

Motivated by \eqref{eq:KSn}, we consider the Cram\'er--von Mises statistic given by
\begin{align}
\mathrm{CvM}_{n,q,\boldsymbol{\gamma}}:=n\int_{-1}^1 \left(F_{n,\boldsymbol{\gamma}}(x)- F_{q}(x)\right)^2\,\mathrm{d}F_q(x).\label{eq:CvMn}
\end{align}
Of course, this statistic still has the issue of being dependent on $\boldsymbol{\gamma}$. Rather than drawing several random directions and aggregating afterwards the tests' outcomes as \eqref{eq:CCFn} does, our statistic itself gathers information from all the directions on $\Omega_{q}$: it is defined as the \textit{expectation} of \eqref{eq:CvMn} with respect to $\boldsymbol{\gamma}\sim\nu_q$:
\begin{align}
\mathrm{CvM}_{n,q}:=\mathbb{E}_{\boldsymbol{\gamma}}\lrc{\mathrm{CvM}_{n,q,\boldsymbol{\gamma}}}=n
\int_{\Omega_q}\lrc{\int_{-1}^1 \lrb{F_{n,\boldsymbol{\gamma}}(x)- F_{q}(x)}^2\,\mathrm{d}F_q(x)}\nu_q(\mathrm{d}\boldsymbol{\gamma}).\label{eq:pcvm}
\end{align}
The test based on \eqref{eq:pcvm} rejects $\mathcal{H}_0$ for large values of $\mathrm{CvM}_{n,q}$.\\

The integration on all possible projection directions within the test statistic, as \eqref{eq:pcvm} does, was firstly considered in the regression context by \cite{Escanciano2006}, though employing an empirical measure instead of $\nu_q$ in \eqref{eq:pcvm}. In our setting, the choice of $\nu_q$ as the distribution of $\boldsymbol{\gamma}$ is canonical, given that it is the only (deterministic) distribution that makes \eqref{eq:pcvm} invariant to rotations of the sample.

\subsection{$U$-statistic form}
\label{sec:Ustat}

Form \eqref{eq:pcvm} is not computationally pleasant: it involves a univariate integral and a more challenging integral on $\Omega_q$. Such level of complexity is undesirable for a test statistic, provided that eventually it may be required to be calibrated by Monte Carlo. In addition, form \eqref{eq:pcvm} obfuscates the quadratic structure of the statistic and complicates obtaining its asymptotic distribution. The next result solves these two issues.

\begin{theo}[$U$-statistic form of $\mathrm{CvM}_{n,q}$;  \cite{Garcia-Portugues:projunif}] \label{th:psi_q:CvM}
	The statistic \eqref{eq:pcvm} can be expressed as
	\begin{align}
	\mathrm{CvM}_{n,q}=\;\frac{2}{n}\sum_{i<j} \psi_q(\cos^{-1}(\mathbf{X}_i'\mathbf{X}_j))+\frac{3-2n}{6},\label{eq:Ustat}
	\end{align}
	where, for $\theta\in[0,\pi]$, 
	\begin{align*}
	\psi_q(\theta) = \begin{cases}
	\frac{1}{2}+\frac{\theta}{2\pi}\left(\frac{\theta}{2\pi}-1\right), & q=1,\\
	\frac{1}{2}-\frac{1}{4}\sin\lrp{\frac{\theta}{2}}, & q=2,\\
	\psi_1(\theta)+\frac{1}{4\pi^2}\lrp{(\pi-\theta)\tan\lrp{\frac{\theta}{2}}-2\sin^2\lrp{\frac{\theta}{2}}}, & q=3,\\
	-\frac{3}{4}+\frac{\theta}{2\pi}+2F_q^2\lrp{\cos\lrp{\tfrac{\theta}{2}}}\\
	\qquad-4\int_0^{\cos\lrp{\theta/2}}F_q(t)F_{q-1}\lrp{\frac{t\tan\lrp{\theta/2}}{(1-t^2)^{1/2}}}\mathrm{d}F_q(t), &q\geq 4.
	\end{cases}
	\end{align*}
\end{theo}

The proof of Theorem \ref{th:psi_q:CvM} is lengthy and therefore omitted. This is also the case for the rest of the presented results. The reader is referred to \cite{Garcia-Portugues:projunif} for the detailed proofs.\\

The case $q=1$ of $\mathrm{CvM}_{n,q}$ is especially interesting. It connects with \cite{Watson1961}'s well-known $U_n^2$ statistic for testing the uniformity of a circular sample, defined as
\begin{align*}
U_n^2:=\,n\int_0^{2\pi}\lrb{F_n(\theta)-F_0(\theta)-\int_0^{2\pi}\lrp{F_n(\varphi)-F_0(\varphi)}\,\mathrm{d}F_0(\varphi)}^2\,\mathrm{d}F_0(\theta),
\end{align*}
where $F_n(\theta):=\tfrac{1}{n}\sum_{i=1}^n1_{\lrb{\Theta_i\leq\theta}}$ is the empirical cdf of the circular sample $\Theta_1,\ldots,\Theta_n$ in $[0,2\pi)$ and $F_0(\theta):=\theta/(2\pi)$ is the uniform cdf on $[0,2\pi)$. The $U_n^2$ statistic can be regarded as the rotation-invariant version of the Cram\'er--von Mises statistic for circular data, achieving such invariance by minimizing the discrepancy of the sample with respect to $\mathcal{H}_0$ (see, e.g., \cite{Garcia-Portugues:review}). \\

The relation between $U_n^2$ and $\mathrm{CvM}_{n,1}$ stems from the following alternative form for $U_n^2$ (see, e.g., \citet[page 111]{Mardia2000}):
\begin{align}
U_n^2= \frac{1}{n}\sum_{i,j=1}^n h\lrp{\Theta_{ij}},\quad  h(\theta):=\frac{1}{2}\lrp{\frac{\theta^2}{4 \pi^2}- \frac{\theta}{2 \pi}+\frac{1}{6}}.\label{eq.Watson.alternative}
\end{align}
Here $\Theta_{ij}:=\cos^{-1}(\cos(\Theta_i-\Theta_j))\in[0,\pi]$ is the shortest angle distance between $\Theta_i$ and $\Theta_j$. Therefore, if we denote by $\Theta_1,\ldots, \Theta_n$ the angles determining the sample $\mathbf{X}_1,\ldots,\mathbf{X}_n$, it happens that $\cos^{-1}(\mathbf{X}_i'\mathbf{X}_j) = \Theta_{ij}$. From this point, elaborating on the expressions in Theorem \ref{th:psi_q:CvM} when $q=1$ leads to that $\mathrm{CvM}_{n,1}=\frac{1}{2}U_n^2$. Therefore our claim that the test based on $\mathrm{CvM}_{n,q}$ is an extension of the Watson test to $\Omega_q$, as stated in the following corollary.

\begin{coro}[An extension of the Watson test to $\Omega_q$] \label{Coro.WatsonEquiv}
	It happens that  $\mathrm{CvM}_{n,1}=\frac{1}{2}U_n^2$. Consequently, the test that rejects for large values of $\mathrm{CvM}_{n,1}$ is equivalent to the Watson test. 
\end{coro}

\subsection{Asymptotic distribution}
\label{sec:asymp}

Expression \eqref{eq:Ustat} unveils the $U$-statistic nature of $\mathrm{CvM}_{n,q}$. Since the $U$-statistic can be seen to be degenerate, the asymptotic distribution of $\mathrm{CvM}_{n,q}$ is an infinite weighted sum of chi-squared random variables. It involves the coefficients $\{b_{k,q}\}$ such that
\begin{align*}
b_{k,q}=\begin{cases}
\frac{2}{\pi}\int_{0}^\pi \psi_1(\theta)T_k(\cos\theta)\,\mathrm{d}\theta,&q = 1,\\
\frac{1}{c_{k,q}}\int_{0}^\pi \psi_q(\theta)C_k^{(q-1)/2}(\cos\theta)\sin^{q-1}(\theta)\,\mathrm{d}\theta,& q\geq 2,
\end{cases}
\end{align*}
where $T_k$ represents the $k$-th Chebyshev polynomial of the first kind, $C_k^{(q-1)/2}$ stands for the $k$-th Gegenbauer polynomial of order $(q-1)/2$, and
\begin{align*}
c_{k,q}:=\displaystyle\frac{2^{3-q}\pi\Gamma(q+k-1)}{(q+2k-1)k!\Gamma((q-1)/2)^2}.
\end{align*}

\begin{theo}[Asymptotic null distribution; \cite{Garcia-Portugues:projunif}]\label{theo:asymp}
	Under $\mathcal{H}_0$ and for $q\geq1$, 
	\begin{align}
	\mathrm{CvM}_{n,q}\stackrel{d}{\rightsquigarrow}\begin{cases}
	\frac{1}{2}\sum_{k=1}^\infty b_{k,1} \chi^2_{d_{k,1}},&q =1,\\
	\sum_{k=1}^\infty \frac{q-1}{q-1+2k} b_{k,q} \chi^2_{d_{k,q}},& q\geq 2,
	\end{cases}\label{eq:asymp}
	\end{align}
	where $\chi^2_{d_{k,q}}$, $k\geq1$, are independent chi-squared random variables with degrees of freedom 
	$$
	d_{k,q}:=\binom{q+k-2}{q-1}+\binom{q+k-1}{q-1}.
	$$
	The coefficients $\{b_{k,q}\}$ are non-negative and satisfy $\sum_{k=1}^\infty b_{k,q} d_{k,q}<\infty$.
\end{theo}

The coefficients $\{b_{k,q}\}$ admit explicit expressions that drastically improve the tractability of the asymptotic null distribution of $\mathrm{CvM}_{n,q}$ for all $q\geq 1$.

\begin{theo}[Coefficients for $\psi_q$; \cite{Garcia-Portugues:projunif}]\label{theo:coefs}
	\label{prop:Gegen_CvM}
	Let $k\geq 1$. For $q\geq1$,
	\begin{align*}
	b_{k,q} = \begin{cases}
	\frac{1}{\pi^2k^2}, & q=1,\\ 
	\frac{1}{2(2k+3)(2k-1)}, & q=2,\\
	\frac{35}{72\pi^2}1_{\{k=1\}} + \frac{1}{2\pi^2}\frac{3k^2+6k+4}{k^2(k+1)(k+2)^2}1_{\{k>1\}}, & q=3,\\
	\frac{(q-1)^2 (2 k+q-1) \Gamma \left((q-1)/2\right)^3 \Gamma \left(3 q/2\right)}{8 \pi q^2 \Gamma \left(q/2\right)^3  \Gamma \left((3 q+1)/2\right)}\\
	\qquad\times{}_4F_3\lrp{1-k,q+k,\tfrac{q+1}{2},\tfrac{3q}{2};q+1,\tfrac{q}{2}+1,\tfrac{3q+1}{2};1},&q\geq 4,
	\end{cases}
	\end{align*}
	where $_4F_3$ stands for the generalized hypergeometric function.
\end{theo}

The final result is a consequence of the fact that $b_{k,q}>0$, for all $k\geq1$ and $q\geq1$, and the fact that $\mathrm{CvM}_{n,q}$ belongs to the class of Sobolev tests \citep{Gine1975}.

\begin{coro}[Omnibusness]
	The test that rejects $\mathcal{H}_0$ for large values of $\mathrm{CvM}_{n,q}$ is consistent against all alternatives to uniformity with square-integrable pdf.
\end{coro}

\section{Numerical experiments}
\label{sec:num}

The asymptotic distributions \eqref{eq:asymp} are usable in practice. The closed forms of $\{b_{k,q}\}$ and the (exact) \cite{Imhof1961}'s method allow to compute asymptotic $p$-values through the evaluation of the truncated-series tail probability function:
\begin{align}
x\mapsto\mathbb{P}\Big[\sum_{k=1}^{K} w_{k,q} \chi^2_{d_{k,q}} > x\Big]\label{eq:tail}
\end{align}
where $x\geq0$ and $K$ is a ``sufficiently large'' integer. Asymptotic critical values $c_\alpha$ for a significance level $\alpha$ are computable using a numerical inversion on \eqref{eq:tail}. \\

The first numerical experiment investigates how large $K$ must be for ensuring a uniform error bound in \eqref{eq:tail}, relatively to $K=10^5$. Figure \ref{fig:1} evidences that \eqref{eq:tail} converges slower, as a function of $K$, for increasing $q$'s. It also gives simple takeaways: (\textit{i}) $K=10^3$ ensures asymptotic $p$-values with uniform error bound $\epsilon=5\times10^{-3}$ for $q\leq10$; (\textit{ii}) $K=10^4$ decreases the uniform error bound to $\epsilon=5\times10^{-4}$; (\textit{iii}) the accuracy for lower $p$-values, approximately in the $[0, 0.15]$-range (left side of the horizontal axis), improves over the uniform bound.\\

The second numerical experiment evaluates the convergence speed of \eqref{eq:asymp} with Table \ref{tab:critval}, which gives the critical values of the statistic for dimensions $q=1,\ldots,10$ and significance levels $\alpha=0.10,0.05,0.01$. As it is unveiled, the convergence towards the asymptotic distribution is quite fast, for all the dimensions explored, effectively requiring to save a single critical value for each dimension $q$ to yield a test decision. The critical values steadily decrease with the increment of the dimension.

\begin{figure}[H]
	\vspace*{-0.25cm}
	\centering
	\includegraphics[width=0.45\textwidth]{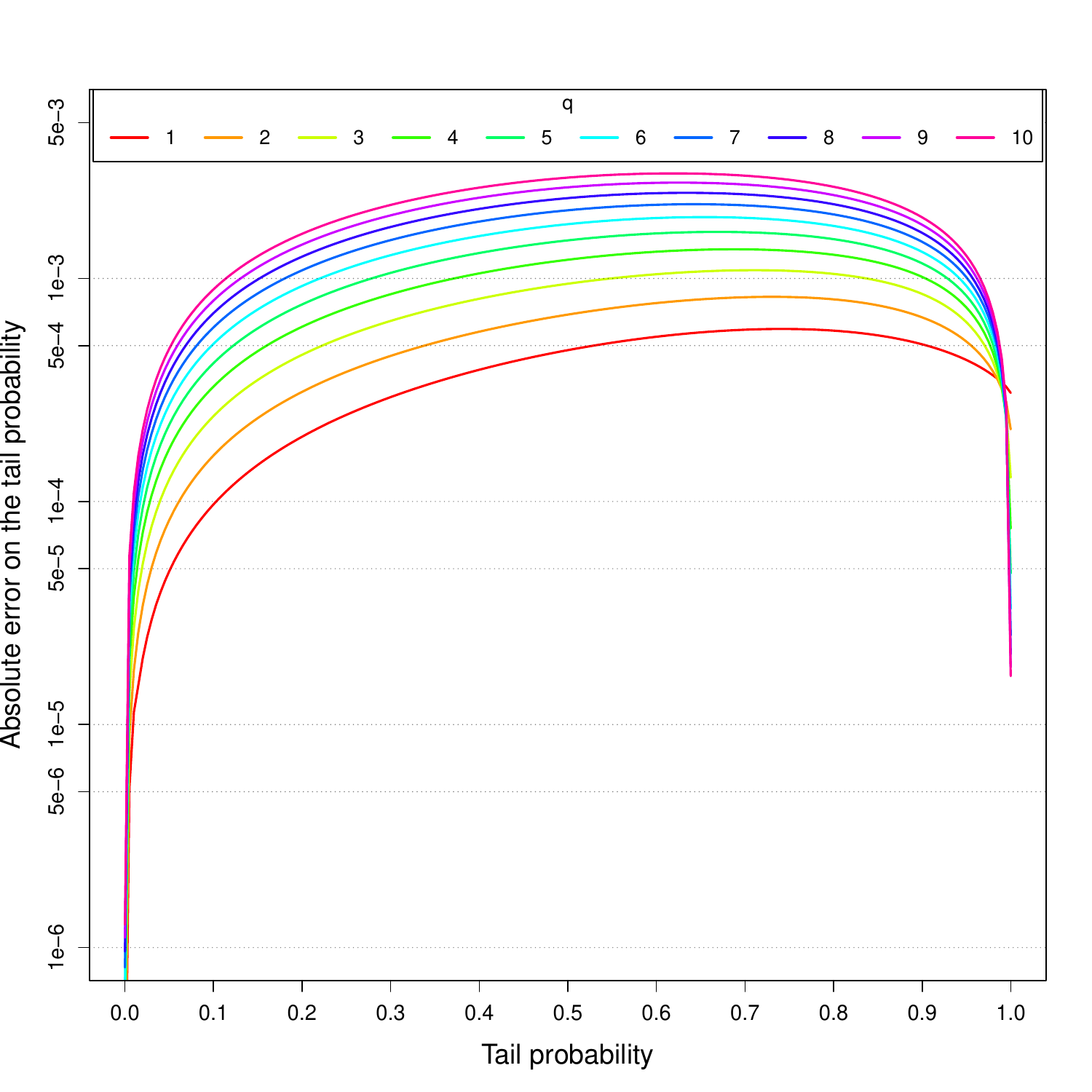}\includegraphics[width=0.45\textwidth]{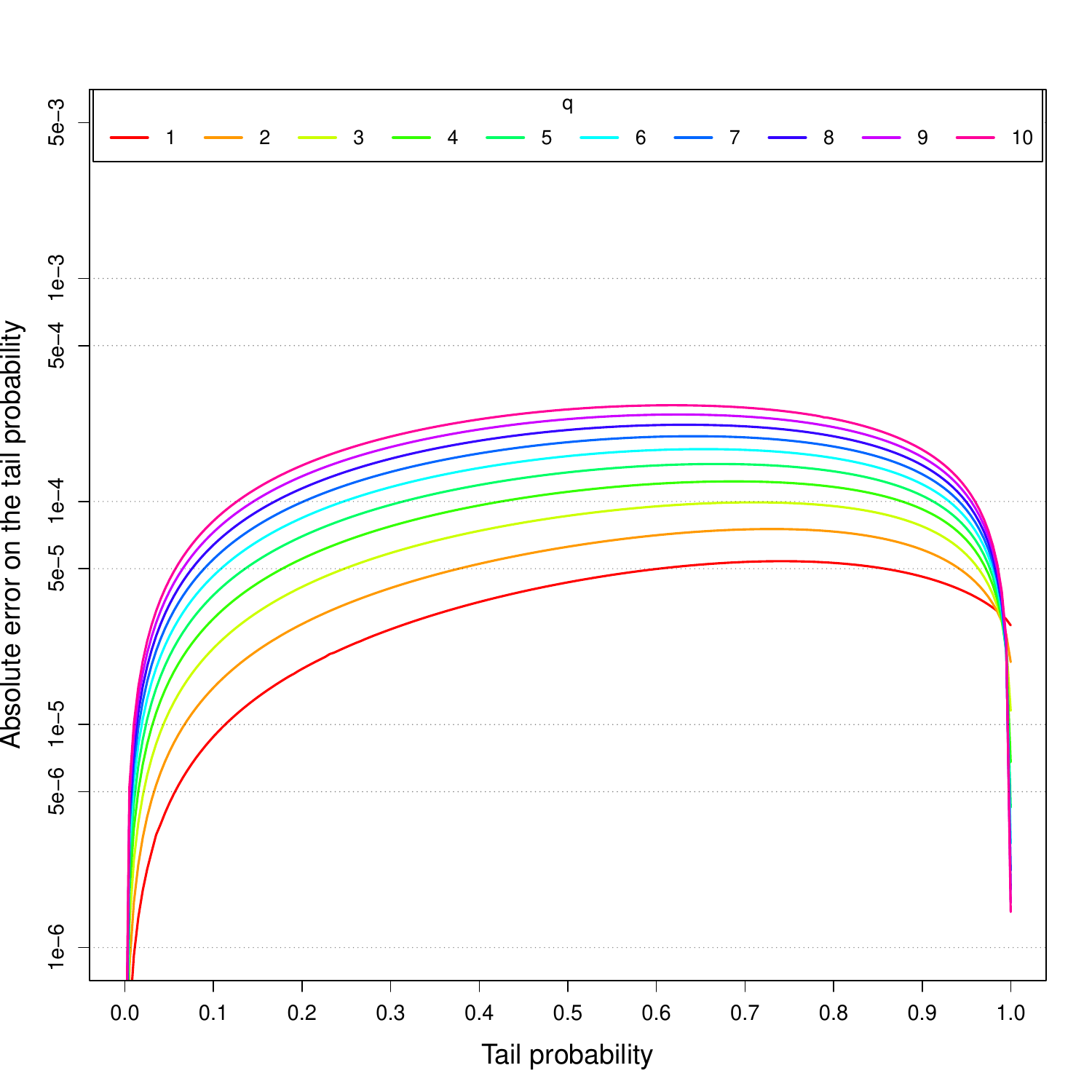}
	\caption{\small Accuracy of the truncation of \eqref{eq:tail}, computed with Imhof's method. The vertical axis shows the absolute errors, with respect to $K=10^5$, of considering $K=10^3$ (left) and $K=10^4$ (right). The horizontal axis shows the probability of \eqref{eq:tail} with $K=10^3$ (a common $[0,1]$-scale for all curves). \label{fig:1}}	
\end{figure}

\begin{table}[H]
	\centering
	\scriptsize
	\newcolumntype{P}[1]{>{\centering\arraybackslash}p{#1}}
	\begin{tabular}{p{0.65cm}p{0.65cm}|P{0.9cm}P{0.9cm}P{0.9cm}P{0.9cm}P{0.9cm}P{0.9cm}P{0.9cm}P{0.9cm}P{0.9cm}P{0.9cm}}
		\toprule
		\multirow{2}{*}{$\alpha$} & \multirow{2}{*}{$n$} & \multicolumn{10}{c}{$q$} \\
		& & $1$ & $2$ & $3$ & $4$ & $5$ & $6$ & $7$ & $8$ & $9$ & $10$ \\ 
		\midrule
		$0.10$ & & & & & & & & & & & \\ 
		& $25$     & $0.3015$ & $0.2752$ & $0.2588$ & $0.2481$ & $0.2401$ & $0.2343$ & $0.2295$ & $0.2256$ & $0.2223$ & $0.2193$ \\
		& $50$     & $0.3026$ & $0.2760$ & $0.2600$ & $0.2490$ & $0.2411$ & $0.2351$ & $0.2302$ & $0.2264$ & $0.2229$ & $0.2201$ \\
		& $100$    & $0.3029$ & $0.2765$ & $0.2605$ & $0.2496$ & $0.2416$ & $0.2355$ & $0.2307$ & $0.2268$ & $0.2234$ & $0.2206$ \\
		& $200$    & $0.3032$ & $0.2769$ & $0.2608$ & $0.2498$ & $0.2419$ & $0.2357$ & $0.2309$ & $0.2270$ & $0.2236$ & $0.2207$ \\
		& $400$    & $0.3036$ & $0.2769$ & $0.2608$ & $0.2502$ & $0.2423$ & $0.2360$ & $0.2311$ & $0.2272$ & $0.2237$ & $0.2209$ \\
		& $\infty$ & $0.3035$ & $0.2769$ & $0.2607$ & $0.2498$ & $0.2419$ & $0.2358$ & $0.2309$ & $0.2269$ & $0.2236$ & $0.2207$ \\
		$0.05$ & & & & & & & & & & & \\ 
		& $25$     & $0.3696$ & $0.3254$ & $0.2994$ & $0.2824$ & $0.2703$ & $0.2613$ & $0.2541$ & $0.2483$ & $0.2434$ & $0.2394$ \\
		& $50$     & $0.3716$ & $0.3273$ & $0.3012$ & $0.2841$ & $0.2719$ & $0.2627$ & $0.2554$ & $0.2495$ & $0.2446$ & $0.2403$ \\
		& $100$    & $0.3730$ & $0.3284$ & $0.3027$ & $0.2852$ & $0.2730$ & $0.2635$ & $0.2563$ & $0.2503$ & $0.2453$ & $0.2411$ \\
		& $200$    & $0.3728$ & $0.3290$ & $0.3029$ & $0.2857$ & $0.2734$ & $0.2638$ & $0.2566$ & $0.2506$ & $0.2456$ & $0.2414$ \\
		& $400$    & $0.3744$ & $0.3288$ & $0.3029$ & $0.2859$ & $0.2735$ & $0.2639$ & $0.2566$ & $0.2508$ & $0.2457$ & $0.2417$ \\
		& $\infty$ & $0.3737$ & $0.3291$ & $0.3029$ & $0.2856$ & $0.2733$ & $0.2639$ & $0.2566$ & $0.2506$ & $0.2456$ & $0.2414$ \\
		$0.01$ & & & & & & & & & & & \\ 
		& $25$     & $0.5220$ & $0.4360$ & $0.3868$ & $0.3561$ & $0.3349$ & $0.3186$ & $0.3062$ & $0.2958$ & $0.2876$ & $0.2805$ \\
		& $50$     & $0.5306$ & $0.4412$ & $0.3920$ & $0.3601$ & $0.3384$ & $0.3219$ & $0.3090$ & $0.2983$ & $0.2903$ & $0.2830$ \\
		& $100$    & $0.5339$ & $0.4451$ & $0.3948$ & $0.3626$ & $0.3400$ & $0.3235$ & $0.3105$ & $0.3002$ & $0.2915$ & $0.2842$ \\
		& $200$    & $0.5359$ & $0.4467$ & $0.3962$ & $0.3642$ & $0.3405$ & $0.3238$ & $0.3112$ & $0.3006$ & $0.2916$ & $0.2843$ \\
		& $400$    & $0.5368$ & $0.4463$ & $0.3968$ & $0.3635$ & $0.3409$ & $0.3242$ & $0.3114$ & $0.3006$ & $0.2921$ & $0.2849$ \\
		& $\infty$ & $0.5368$ & $0.4469$ & $0.3963$ & $0.3639$ & $0.3413$ & $0.3244$ & $0.3113$ & $0.3008$ & $0.2921$ & $0.2848$ \\
		\bottomrule
	\end{tabular}
	\caption{\small Critical values of the $\mathrm{CvM}_{n,q}$ statistic, approximated by $M=10^6$ Monte Carlo replicates. The asymptotic ($\infty$) critical values result from computing and inverting \eqref{eq:tail} with $K=10^4$. \label{tab:critval}}
\end{table}

\section{Are Venusian craters uniformly distributed?}
\label{sec:venus}

Venus is the closest planet to Earth and the most Earth-like planet of the Solar System in terms of size and composition. As such, it is one of the most explored extraterrestrial bodies by humankind, a landmark on its exploration being the Magellan mission (1989--1994). Through a series of mapping cycles, the Magellan spacecraft produced the first global, high-resolution mapping of $98\%$ of the Venusian surface. The analysis of the vast imagery produced in the mission (see \cite{Ford1993}) revealed the high uniformity of the Venusian craters distribution \citep{Schaber1992,Phillips1992}. Indeed, \citet[Section 3.1]{Phillips1992} tested the uniformity of such distribution using the $763$ locations of craters back then available, finding no evidence to reject $\mathcal{H}_0$ for several tests. \\

We tested uniformity with an updated database of Venusian craters \cite{VenusData} that contains the locations of the centres of $967$ craters. Figure \ref{fig:2} shows these locations over Venus surface, as mapped by the Magellan mission. We performed the \cite{Rayleigh1919} and \cite{Gine1975}'s $F_n$ tests, as considered by \cite{Schaber1992}, the \cite{Cuesta-Albertos2009} test (using $k=50$), and the novel $\mathrm{CvM}_{n,2}$-based test. The obtained $p$-values, estimated with $10^4$ Monte Carlo replicates, were $0.170$, $0.112$, $0.117$, and $0.129$, respectively. Consequently, we found no statistical evidence to reject $\mathcal{H}_0$ at usual significance levels, thus confirming the analysis by \cite{Phillips1992} with updated crater records.\\

The apparent uniform distribution of Venusian craters is truly remarkable. Indeed, among the very few planets and moons of the Solar System with uniformly-distributed craters, Venus has the largest number of observed craters, according to the database of named craters of the International Astronomical Union \citep{Garcia-Portugues:projunif}. The filtering of small meteoroids by the dense Venusian atmosphere may be one of the causes explaining the uniform distribution of craters.

\begin{figure}[h]
	\centering
	\includegraphics[width=0.9\textwidth]{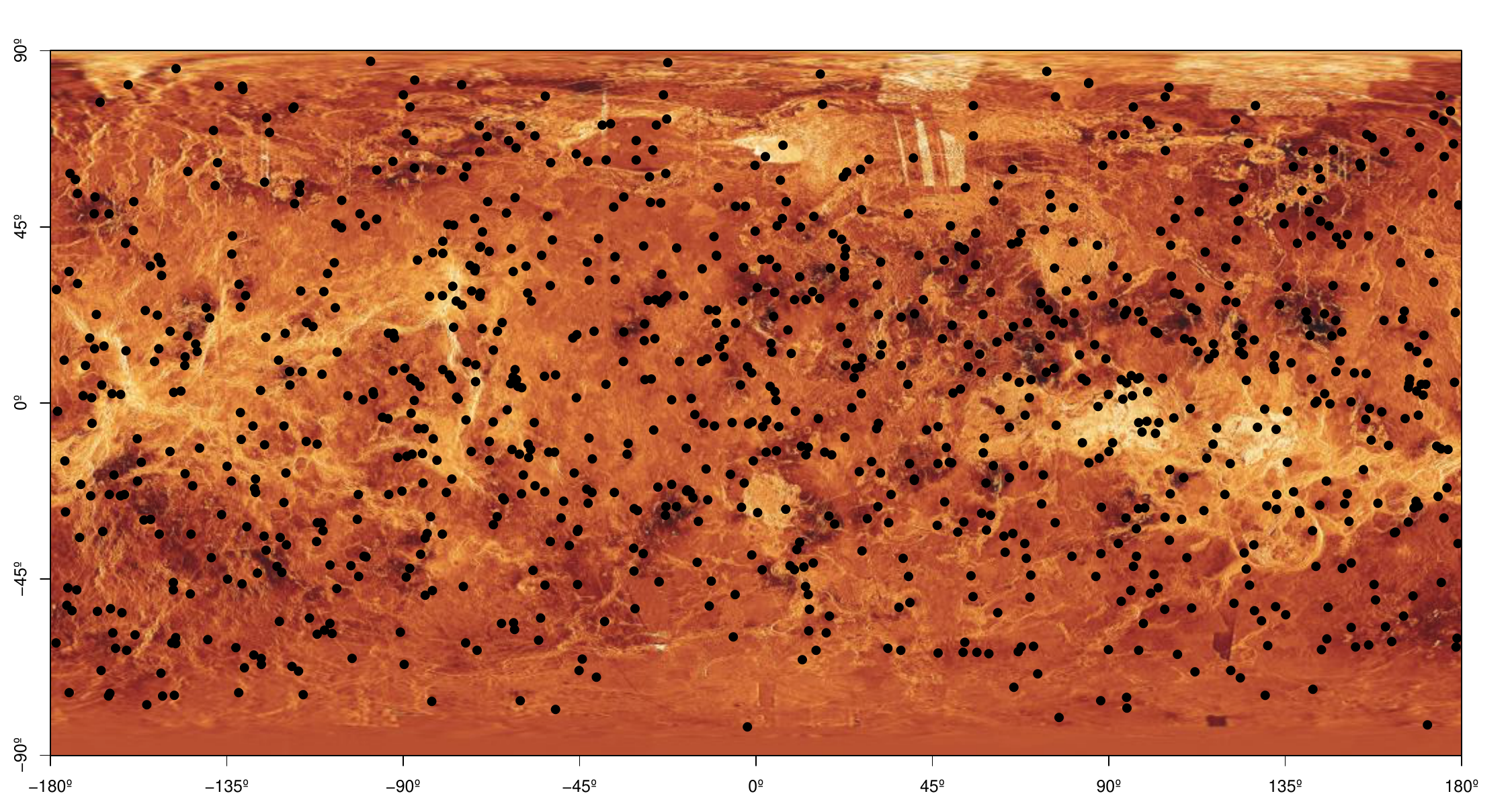}
	\caption{\small Venusian craters (black points) overlaid over a colourized image of the Venus surface \citep{VenusImage}.\label{fig:2}}	
\end{figure}

\section*{Acknowledgements}

The first author acknowledges financial support from grants PGC2018-097284-B-I00, IJCI-2017-32005 and MTM2016-76969-P, funded by the Spanish Ministry of Economy, Industry and Competitiveness, and the European Regional Development Fund. The second and third authors acknowledge financial support from grant MTM2017-86061-C2-2-P from the Spanish Ministry of Economy, Industry and Competitiveness. The authors gratefully acknowledge the computing resources of the Supercomputing Center of Galicia (CESGA). Comments by two referees are acknowledged.


\end{document}